\documentclass[12pt]{article}         
\usepackage{epsfig}                   
\newcommand{\h}{\hspace*{5 ex}}
\newfont{\ninerm}{cmr9}
\def\INFN{ Istituto Nazionale di Fisica Nucleare, Sezione di Firenze}
\def\Zipcode{I-50125 Firenze, Italy}
\def\Dipa{ Dipartimento di Fisica, Universit\`a di Firenze}
\def\dfrac{\displaystyle\frac}
\def\alfa{4\pi\alpha}
\def\g2r2{\frac{G}{2\sqrt{2}}}

\def\geff{g_{eff}}
\def\gve{g_V^e}
\def\gae{g_A^e}
\def\GE{G_E}
\def\GM{G_M}
\def\tGE{\tilde{G}_E}
\def\tGM{\tilde{G}_M}
\def\tGA{\tilde{G}_A}
\def\GAS{G_A^{(s)}(Q^2)}
\def\GES{G_E^{(s)}(Q^2)}
\def\GMS{G_M^{(s)}(Q^2)}
\def\gaesse{g_A^{(s)}}

\def\mus{\mu_s}

\def\rdues{r^2_s}
\def\RAT1{R_A^{T=1}}
\def\RATZERO{R_A^{T=0}}
\def\RAS{R_A^{(0)}}
\def\laes{\lambda_E^{(s)}}
\def\lams{\lambda_M^{(s)}}
\def\laas{\lambda_A^{(s)}}
\def\j5{j_5}
\def\js5{j_5^+}
\def\J#1#2#3{J^{#1#2}_{#3}}
\def\em{[\gamma]}
\def\emV{[\gamma V]}
\def\emA{[\gamma A]}
\def\Ze{[Z]}
\def\rad2{\sqrt{2}}
\def\rtuptau{\sqrt{\tau(1+\tau)}}
\def\rumep2{\sqrt{1-\varepsilon^2}}
\def\rtepumep{\sqrt{\tau\varepsilon(1-\varepsilon)}}
\def\ruptepupep{\sqrt{(1+\tau)\varepsilon(1+\varepsilon)}}
\def\CSEN{ \dfrac{d\sigma}{d\Omega_{e^\prime}} }
\def\CLAB{ \dfrac{\alpha^2}{Q^4} \Biggr(\dfrac{E_{e^\prime}}{E_e}\Biggr)^2 }
\def\SMEEP{ \sigma_M \frac{E_{e^\prime}}{E_e} }
\def\sigmas{ \sigma(s) }
\def\sigmaszero{ \sigma(0) }
\def\Dsigmas{ \Delta\sigmas }

\def\Ax{ {\cal A}_x }
\def\Az{ {\cal A}_z }
\def\Ae{ {\cal A}_{LR} }
\def\tep{ \vartheta_{e'} }
\def\tg2{\tan^2{ \frac {\vartheta_{e'}}{2} } }
\def\t2g{\tan^2(\vartheta_{e'}/2)}
\def\c2t{\cos^2(\vartheta_{e'}/2)}
\def\s4t{\sin^4(\vartheta_{e'}/2)}
\def\tkq{\vartheta_{\hat{kq}}}
\def\L#1#2{L^{#1}_{#2}}
\def\L#1#2{L^{#1}_{#2}}
\def\W#1#2#3{W^{#1#2}_{#3}}
\title{
~~~~~~~~~~~~~~~~~~~~~~~~~~~~~~~~~~~~~~~~~~~DFF 330-11-98  \\
\vspace{2. cm}
{\bf Parity violating target asymmetry \\
     in electron - proton scattering} }

\author{M. Moscani, B. Mosconi \\                    
                {\small  \Dipa } \\
                {\small  \INFN } \\
                {\small  \Zipcode}   \\
and \\
P. Ricci\\
                {\small  \INFN } \\
                {\small  \Zipcode }
}
\date{\today}                        
\begin{document}
\baselineskip=20pt
\maketitle
\thispagestyle{empty}

\begin{abstract}
We analyze the parity-violating (PV) components of the analyzing power 
${\cal A}$ in elastic electron-proton scattering and discuss their 
sensitivity to the strange quark contributions to the proton weak form 
factors . 
We point out that the component of ${\cal A}$ along the momentum 
transfer is independent of the electric weak form factor and thus 
compares favorably with the PV beam asymmetry for a determination of the 
strangeness magnetic moment. 
We also show that the transverse component could be used for 
constraining the strangeness radius. 
Finally, we argue that a measurement of both components could give 
experimental information on the strangeness axial charge. 
\end{abstract}
\
\
PACS numbers:  13.10.+q , 13.40.Gp , 13.85.Dz , 14.20.Dh 
\newpage
\section{Introduction}
\par
Parity violating (PV) electron scattering 
can provide very interesting information on the electroweak (ewk)
structure of the nucleon, as first suggested in Ref.\cite{McKeown89,Beck89}, 
complementary to those given by neutrino scattering and by PV atomic 
experiments. 
\par
Only three of the four form factors determining the proton weak neutral 
currents (excluding second class currents) can be probed by the PV electron 
scattering. In fact, the induced pseudoscalar 
component of the axial-vector current does not contribute to PV electron 
scattering to leading order in ewk coupling. 
\par
In particular it could shed light on the possible strange quark 
contributions to the nucleon properties. Such a contribution is strongly 
suggested by the analysis of the pion-nucleon sigma term 
\cite{Cheng76,DoNa86,GaLe91}, of the cross section of the elastic 
neutrino/antineutrino-proton scattering \cite{Ahrens87,Garvey93}, and of 
the spin-dependent structure functions of the nucleon in deep inelastic 
scattering \cite{Ashman8889,Adams94,Abe95,Adeva9394}. 
\par
A first measurement of the PV beam asymmetry in ${\vec e}-p$ elastic 
scattering was performed at Bates/MIT Laboratory by the SAMPLE 
Collaboration giving the first experimental determination of the 
weak magnetic form factor of the proton at Q$^2$= 0.1 (GeV/c)$^2$ 
\cite{Mue97}.  
From this value the strangeness magnetic form factor is obtained 
by subtracting out the neutron and proton magnetic form factors as 
G$_M^s$(0.1 (GeV/c)$^2$) = 0.23 $\pm$ 0.37 $\pm$ 0.15 $\pm$ 0.19 $\mu_N$, 
where the three uncertainties are statistical, systematic and 
theoretical, respectively. 
\par
This result suggests a positive value of the strangeness magnetic 
moment $\mu_s$ = G$_M^s$(0), even if a vanishing or a negative value can 
not be excluded. This is at variance 
with the large majority of the theoretical predictions and could lead 
to rather bizarre properties of QCD \cite{Leinw98}. 
\par
A second measurement of the PV beam asymmetry at $Q^2$=0.48 (GeV/c)$^2$ 
has been made very recently by the HAPPEX Collaboration \cite{HAPPEX98}~  
at the Thomas Jefferson Laboratory. Unlike the SAMPLE Collaboration, they 
have chosen to detect the scattered electrons at forward angles 
so as to provide information on the strange contribution 
to the weak electric form factor $\tGE$. The measured asymmetry 
($\Ae$ = - 14.5 $\pm$ 2.2 ppm) turns out to be explained by the ewk Standard 
Model without contributions from the strange quark. 
\par
Because of the difficulties inherent in the PV electron scattering experiment
an independent determination of $\mu_s$, as well as of the other
strangeness properties of the proton (mean-square radius and axial charge), 
could be extremely useful. 
\par
The literature is already rich of several theoretical studies 
\cite{Beck89,MusolfPRC94,Donnelly92,Horowitz93,Amaro96} 
and experimental proposals \cite{Bates94,Mainz93,TJNAF1,TJNAF2} 
~of PV electron scattering from complex nuclei. 
In the case of the deuteron, the PV helicity asymmetry 
of the exclusive electrodisintegration cross section has also been 
considered \cite{MR97}. 
\par
Another opportunity not yet explored is given by the measurement 
of the PV asymmetry of the analyzing power arising in the scattering 
of unpolarized electrons from polarized protons. In principle, the PV 
target asymmetry is even more versatile than the PV 
beam asymmetry for disentangling the different weak form factors 
because the polarization of the proton target can be freely chosen 
whereas the electron beam can be polarized only along the beam momentum. 
\par
In this paper we intend to study such target asymmetry in the low 
Q$^2$ range and to analyze its potential for an experimental 
determination of the ewk form factors. 
\par
The rapid and continuous improvements in the production of polarized 
targets \cite{Haeberli97}~ promise to make feasible in the near future 
such a measurement. In fact, experiments with polarized hydrogen targets 
are nowadays very common either using solid targets of hydrogenous compounds 
such as NH$_3$ or Butanol with external beams or gas targets  
with internal beams. 
High proton spin polarization ($\sim 90\%$) can be achieved in solid targets 
using the method of dynamic nuclear polarization 
together with the $^3$He/$^4$He dilution refrigerator technique. 
Also the polarization reversal is now a relatively fast process 
with high spin flip efficiency. Further, the production of polarized internal 
targets in storage rings seems to be very promising because it allows one to 
have pure atomic species, high polarization and polarization reversibility  
on a time scale of msec. Moreover, high thickness can be reached 
thanks to the laser optical pumping, together with 
a free choice of the spin orientation using Helmholtz coils. 
In conclusion, a measurement of the PV analyzing power in the $e-{\vec p}$ 
scattering seems within the reach of the present experimental capability, 
provided the entire apparatus is kept stable for the long time required. 
\par
Note that other more complicated PV observables are possible in the electron 
proton elastic scattering if both beam and target are polarized. 
However, they are out of the experimental feasibility at the moment, and 
thus we shall not consider them in this paper. 
\par
Alternatively to the measurement of the PV asymmetry in the analyzing 
power, the PV polarization of the recoiling proton for 
unpolarized beam and target could be studied \cite{Moscani97}. However, 
even if the recoil polarimetry has developed rapidly in recent years, 
the PV longitudinal and sideways components
of the polarization vector have values which are typically 
of the order of 10$^{-5}$ at low Q$^2$, i.e. 
three-order of magnitude smaller than those measurable with the 
actual focal plane polarimeters. 
\par
A work on target asymmetry has been published some years ago by Fayyazuddin 
and Riazuddin \cite{FR90}~ who limited their considerations to the 
longitudinal component of the analyzing power. 
However, their results are useful for the order of magnitude, at the most, 
because their analytic expression  contains several errors (an overall 
factor of 2, a sign error and an erroneous dependence on the form factors 
in the eletromagnetic (em) - axial interference term). 
This is rather strange because the effects of the neutral currents in the 
elastic electron-proton scattering had already been considered by different 
authors in the 70's, a few years after the appearance of the unified 
gauge theory of weak and em interactions \cite{Weinberg67,Salam68}. 
While much work was concentrated on the polarized beam asymmetry, 
in \cite{CNg77,RS74,BDC76}~ the case of the polarized target asymmetry was 
also considered. In these papers the analytic expression of the PV target 
asymmetry is incorrect only in the sign of the em - axial interference term 
\cite{CNg77,RS74}~ or in the sign of the em - vector interference term 
\cite{BDC76}. Note that these papers were mainly concerned 
with the discrimination among various SU(2)X U(1) models 
and other models based on larger gauge groups, such as SU(2)$_L$ X SU(2)$_R$ 
or SU(3) X U(1). 
\par
Finally, we have to mention that both PV beam and target asymmetries in 
the electron-proton scattering have also been investigated \cite{Hwang88}~ 
in the GeV range to detect manifestations of a possible nonstandard weak 
boson $Z^{\prime 0}$. In this paper analytic formulas for the asymmetries 
are not reported because they are evaluated in a numerical way starting 
from the ewk currents. 
\par
The rest of the paper is organized as follows. In Sec.2 we describe our 
formulation of the elastic $e-{\vec p}$ scattering and we give the 
expressions of the target asymmetries. In Sec.3 we present and discuss 
our numerical results. Finally, in Sec.4 we state our conclusions. 
\section{Formalism}
\subsection{Parity-violating elastic cross section}
\h The invariant amplitude for 
the parity-violating elastic electron proton scattering 
to lowest order, is the sum of the 
one-photon and the one-$Z^0$ boson exchange process: 
%
%
\begin{equation}
  {\cal{M}} = {\cal{M}}_{\em} + {\cal{M}}_{\Ze} ~~~~~~,\label{e:M} 
\end{equation}
with
\begin{equation}
 {\cal{M}}_{\em} = -\dfrac{\alfa}{Q^2} ~j_\mu ~\J{\em}{\mu}{} 
                                         ~~~~~~, \label{e:Mgamma}
\end{equation}
and
\begin{equation}
  {\cal{M}}_{\Ze} = \g2r2  (\gve j_\mu + \gae  j_{\mu 5})
                      ~\J{[NC]}{\mu}{} ~~~~~~, \label{e:Mzeta}
\end{equation}
in the limit $Q^2 \ll M_Z^2$, which we are interested in; 
$Q^2 = - q_\mu^2 > 0$ is the four momentum transfer squared, $\alpha$ is 
the fine-structure constant, $G$ is the weak Fermi constant. Finally, 
$\gve$ and $\gae$ are the neutral vector and axial-vector couplings of the 
electron which, in the Standard Model, are given by $\gae = 1$, 
$\gve = -1+4 \sin^2\vartheta_W$, $\vartheta_W$ being the Weinberg or 
weak-mixing angle. The conventions of Musolf et al. \cite{MuDo94}~ for the 
weak coupling constants are assumed. 
\par
The electron vector and axial-vector currents are given by the Dirac form
%
%
\begin{eqnarray}
          j_\mu &=& \bar u(k^\prime,s^\prime_e) \gamma_\mu u(k,s_e) 
                                   ~~~~~,\nonumber \\
      j_{\mu 5} &=& \bar u(k^\prime,s^\prime_e) \gamma_\mu \gamma_5 u(k,s_e) 
                                   ~~~~~~, \label{e:jmu}
\end{eqnarray}
where $(k,s_e)$ and $(k^\prime,s^\prime_e)$ are the four-momentum and the 
covariant spin four-vector of the incoming and outgoing electron, 
respectively. We recall the properties  
$k_e\cdot s_e = k_e^\prime \cdot s_e^\prime = 0$ 
and $s_e^2 = s_e^{\prime 2}=-1$. 
\par
As for the nucleonic currents, $\J{\em}{}{\mu}$ is the em 
current and $\J{[NC]}{}{\mu}$ the neutral current which consists of a 
vector and an axial-vector component
%
%
\begin{equation}
       \J{[NC]}{}{\mu} = \J{[V]}{}{\mu} + \J{[A]}{}{\mu} ~~~~~~~.\label{e:JNC}
\end{equation}
\par
Using the Gordon decomposition, the general 
expressions of the matrix elements of the nucleon ewk currents 
consistent with Lorentz covariance and with parity and time-reversal 
invariance can be written in the useful form 
%
%
\begin{eqnarray}
  \J{\em}{}{\mu}  &=& \bar u(p^\prime,s^\prime) 
                \Biggr[G_M \gamma_\mu + \dfrac{(G_E - G_M)}{(1+\tau)} 
                         \dfrac{(p+p^\prime)_\mu}{2M} \Biggr] u(p,s) 
                                               ~~~~,\nonumber \\ 
  \J{[V]}{}{\mu} &=& \bar u(p^\prime,s^\prime) 
                \Biggr[\tilde G_M \gamma_\mu 
              + \dfrac{(\tilde G_E - \tilde G_M)}{(1+\tau)} 
                \dfrac{(p+p^\prime)_\mu}{2M} \Biggr] u(p,s) 
                                                ~~~~,\label{e:Jemnc} \\ 
  \J{[A]}{}{\mu} &=& \bar u(p^\prime,s^\prime) \Big[\tilde G_A \gamma_\mu  
            + i(\tilde G_P/M) q_\mu \Big] \gamma_5 u(p,s) 
                                                ~~~~,\nonumber 
\end{eqnarray}
where $M$ is the nucleon mass, $\tau = Q^2/4M^2$. As before, the proton 
four-momenta and spin four-vectors satisfy the relations 
$p\cdot s = p^\prime \cdot s^\prime = 0$ and $s^2 = s^{\prime 2}=-1$. 
\par
In the following, we do not need to care about the induced pseudoscalar 
current which, being proportional to $\gamma_5 q_\mu$, does not contribute 
to observables in PV electron scattering to leading order in ewk coupling 
when saturated with the leptonic tensor. 
\par
Assuming, as usual, the extreme relativistic limit (ERL) for the electron 
($m_e \ll E_e$) and neglecting the purely weak component 
term, which is $\sim G^2$, the elastic cross section in the case of an 
unpolarized electron beam and a polarized nucleon target, can be written
%
%
\begin{eqnarray}
 \CSEN &=& \CLAB  ~\sum_{s^\prime}
              ~\Biggr\{ \L{0}{\mu\nu} \W{}{\mu\nu}{\em}   
                                                             \nonumber \\
      &-& \geff ~\Big(\gve \L{0}{\mu\nu} + \gae  \L{h}{\mu\nu}\Big) 
          \Big( \W{}{\mu\nu}{\emV} + \W{}{\mu\nu}{\emA} \Big)\Biggr\}
                           ~~, \label{e:csen} 
\end{eqnarray}
where we have averaged over the initial electron spin states 
and summed over the final ones. 
The effective weak coupling constant $\geff$ determining the magnitude of 
the PV effects in the low and medium $Q^2$ is given by 
%
%
\begin{equation}
 \geff= {\frac{Q^2}{4\pi\alpha}} {\frac{G}{2 \sqrt{2}}}
            ~~~~,\label{e:geff}
\end{equation}
and
%
%
\begin{eqnarray}
    \L{0}{\mu\nu} &=& 2 \Big( k_\mu k^\prime_\nu + k_\nu k^\prime_\mu 
                           - \frac{Q^2}{2} g_{\mu\nu} \Big)
                              ~~~~~,\nonumber \\
    \L{h}{\mu\nu} &=& 2 ~i ~\epsilon_{\mu\nu\alpha\beta} q^\alpha k^\beta 
                              ~~~~~~, \label{e:L0hmunu}
\end{eqnarray} 
are the well known symmetric and antisymmetric 
leptonic tensors, where $\epsilon_{\mu\nu\alpha\beta}$ 
is the totally antisymmetric Levi-Civita tensor. 
\par
Exploiting Lorentz and CP invariance (but allowing for parity violation), 
the hadronic tensor can be expressed in the general form ~\cite{Ans95} 
%
%
\begin{eqnarray}
 \W{}{\mu\nu}{[a]} &=& F^{[a]}_1(Q^2) g^{\mu\nu} 
                  + F^{[a]}_2(Q^2) \frac{p^\mu p^\nu}{M^2} ~~~~~ \nonumber \\
                 &+& i \epsilon^{\mu\nu\alpha\beta} \Biggr[ 
                    F^{[a]}_3(Q^2) \frac{p_\alpha q_\beta}{M^2} 
                  + g^{[a]}_1(Q^2) \frac{q_\alpha s_\beta}{M} 
                  + g^{[a]}_2(Q^2) \frac{p_\alpha s_\beta}{M} \Biggr] 
                                                         \label{e:W0s2} \\
                 &+& g^{[a]}_3(Q^2) \frac{p^\mu s^\nu + p^\nu s^\mu}{M} 
                  + g^{[a]}_4(Q^2) \frac{(q\cdot s) p^\mu p^\nu}{M^3} 
                  + g^{[a]}_5(Q^2) \frac{q\cdot s}{M} g^{\mu\nu} 
                                                            ~~,  \nonumber
\end{eqnarray}
where $a=\gamma, \gamma V, \gamma A$.
In the case of the elastic scattering, the structure functions of the 
deep inelastic scattering case considered in \cite{Ans95}, 
become the unpolarized form factors $F^{[a]}_i$ and the polarized ones 
$g^{[a]}_i$, while the structure of the tensor remains unchanged. 
\par
Note that in the literature are present other different (but equivalent) 
definitions of the hadronic tensor and of the structure functions 
obtained from each other by inserting or dropping terms proportional to 
$q^\mu$ or $q^\nu$ (which give no contribution in the ERL) 
and using suitable identities ~\cite{Ans95,Hei73,Bartelski79}. 
\par
From the explicit calculation the non-zero form factors are 
%
%
\begin{eqnarray}
 F^{\em}_1  &=& - ~\tau ~\GM^2  ~~~~,\nonumber \\
 F^{\em}_2  &=&   \frac{\GE^2 + \tau \GM^2}{1+\tau} ~~~~,\nonumber \\
 g^{\em}_1  &=& - ~\frac{\GM(\GE + \tau \GM)}{2(1+\tau)} ~~~~,\nonumber \\
 g^{\em}_2  &=&   \tau ~\frac{\GM(\GE - \GM)}{1+\tau} ~~~~,\nonumber \\
 F^{\emV}_1 &=& - ~2 ~\tau ~\GM \tGM ~~~~,\nonumber \\
 F^{\emV}_2 &=&   2 ~\frac{\GE \tGE + \tau \GM \tGM}{1+\tau} 
                                                     ~~~~,\label{e:Fg} \\
 g^{\emV}_1 &=& - ~\frac{\GM \tGE + \GE \tGM + 2 \tau \GM \tGM}{2(1+\tau)}
                                                     ~~~~,\nonumber \\
 g^{\emV}_2 &=&   \tau ~\frac{\GM \tGE + \GE \tGM - 2 \GM \tGM}{1+\tau}
                                                     ~~~~,\nonumber \\
 F^{\emA}_3 &=&   \tGA \GM                          ~~~~,\nonumber \\
 g^{\emA}_3 &=&   \tGA \GE                          ~~~~,\nonumber \\
 g^{\emA}_4 &=& - ~\frac{\tGA(\GE - \GM)}{1 + \tau}  ~~~~,\nonumber \\
 g^{\emA}_5 &=& - ~\tGA \GM                          ~~~~.\nonumber
\end{eqnarray}
\par
Let us introduce the notation $\sigmas \equiv (d\sigma/d\Omega_{e^\prime})$ 
to indicate the elastic cross section. 
As it is clear from the general structure of the hadronic tensor, 
Eq.(\ref{e:W0s2}), $\sigmas$ can be written 
%
%
\begin{equation}
  \sigmas = \sigmaszero + \Dsigmas
            ~~~~~,\label{e:sigmas}
\end{equation}
where $\Dsigmas$ is the contribution deriving from the initial nucleon 
polarization as obtained from the corresponding part of the hadronic 
tensors $\W{}{\mu\nu}{[a]}$. 
\par
From the symmetry properties in $\mu \leftrightarrow \nu$ of the leptonic 
and hadronic tensors,the cross section, Eq.(\ref{e:csen}), reads
%
%
\begin{eqnarray}
 \sigmaszero 
      &=& \CLAB \Biggr\{ \L{0}{\mu\nu} \W{}{\mu\nu}{\em}(S)    
                                                 \label{e:sigmaszero} \\
      &-& \geff \Biggr( \gae  \L{h}{\mu\nu} \W{}{\mu\nu}{\emA}(A) 
       + \gve  \L{0}{\mu\nu} \W{}{\mu\nu}{\emV}(S) \Biggr) \Biggr\}
                                                ~~~~, \nonumber
\end{eqnarray}
%
%
%
%
\begin{equation}
 \Dsigmas = - \CLAB ~\geff \Biggr( \gae  \L{h}{\mu\nu} 
                                         \W{}{\mu\nu}{\emV}(A) 
            + \gve  \L{0}{\mu\nu} \W{}{\mu\nu}{\emA}(S) \Biggr) 
                                           ~~~~,\label{e:Deltasigmas} 
\end{equation}
where $\W{}{\mu\nu}{[a]}(S)$ and $\W{}{\mu\nu}{[a]}(A)$ are the 
symmetric and antisymmetric components of the hadronic tensor. 
In the laboratory (lab) frame ($p^\mu = (M,{\bf 0})$) the 
cross section becomes
%
%
\begin{eqnarray}
\sigmaszero &=&
    \SMEEP \frac{1}{\varepsilon(1+\tau)} 
             \Biggr\{\Big(\varepsilon~ G_E^2 + \tau G_M^2\Big) \nonumber \\
             &-&2 ~\geff \Biggr[ \gve  \Big( \varepsilon~ G_E {\tilde G}_E 
                             + \tau G_M {\tilde G}_M  \Big) 
                                                      \label{e:sigmaszero1}\\
             &+& \gae  \sqrt{\tau(1+\tau)(1-\varepsilon^2)} 
                                  G_M {\tilde G}_A  \Biggr] \Biggr\} 
                                                           ~~~~, \nonumber 
\end{eqnarray}
%
%
%
%
\begin{eqnarray}
 \Dsigmas &=& - ~\geff ~\SMEEP ~\dfrac{1}{\varepsilon(1+ \tau)} 
                                                                 \nonumber \\
          &\times& ~\Biggr\{  \gae ~\Biggr(\dfrac{1-\varepsilon}{M}\Biggr) 
                 \Biggr[ (G_E \tGM + \tGE G_M) (k\cdot s) 
                                                                 \nonumber \\
      &-& \Biggr( (G_E \tGM + \tGE G_M)(E_e+M) - 2 G_M\tGM (E_e-M\tau) \Biggr) 
                  \dfrac{(q\cdot s)}{2M(1+\tau)} \Biggr]
                                                                 \nonumber \\
      &+& ~\gve ~\Biggr(\dfrac{1}{2M^2 \tau}\Biggr)
          ~\tGA ~\Biggr[ 2 G_E (1-\varepsilon) (E_e-M\tau) (k\cdot s)  
                                                    \label{e:Dsigmas2}\\
      &+&  \Biggr( 2 M \tau G_M 
                 - G_E \Big( E_e(1-\varepsilon) + 2 M \tau \varepsilon \Big)  
                                   \Biggr)(q\cdot s) \Biggr] \Biggr\} 
                                                                 ~, \nonumber
\end{eqnarray}
where
%
%
\begin{equation}
 \sigma_M = \frac{\alpha^2 \c2t}{4 E^2_e \s4t } 
             ~~~~~~, \label{e:Mott} 
\end{equation}
is the Mott cross section for elastic scattering of an electron 
from a structureless proton and the parameter $\varepsilon$, which gives 
the degree of transverse linear polarization of the virtual 
photon, is defined as
%
%
\begin{equation}
 \varepsilon = \Big[1 + 2 (1 + \tau) \t2g\Big]^{-1} 
                                    ~~~~. \label{e:epsilon1}
\end{equation}
\par
The cross section asymmetry due to the target polarization is defined as 
%
%
\begin{equation}
  {\cal A}_s = \frac{\sigma(s) - \sigma(-s)} 
                                    {\sigma(s) + \sigma(-s)} 
                             ~~~~~,\label{e:Apt1}
\end{equation}
where $\sigma(\pm s)$ is the cross section when the initial proton 
is polarized along ($s$) or opposite ($-s$) an arbitrary direction 
${\hat {\bf s}}$. Then, from Eq.(\ref{e:sigmas}), we see that 
$\Big( \sigma(s) - \sigma(-s) \Big) = 2 ~\Dsigmas$
and only the interference terms contribute to the asymmetry; moreover 
$\Big(\sigma(s) + \sigma(-s)\Big) = 
2 ~\sigmaszero \simeq 2 ~\sigma_{\em}(0)$ because the contribution deriving 
from the em-weak interference can be disregarded.
\par
As for the reference frame we take, as usual, the y-axis orthogonal 
to the scattering plane 
(${\hat {\bf e}}_y ~// ~{\bf k} \times {\bf k^\prime}$), 
while for the direction of the z-axis there are substantially two possible 
choices: z-axis along the direction {\bf k} of the 
incident beam or along the direction {\bf q} of the momentum 
transfer. In both cases, obviously, the only nonzero components of the 
asymmetry are those in the scattering plane, i.e. the transverse 
(${\cal A}_x$) and the longitudinal (${\cal A}_z$) ones. 
\par
From the theoretical point of view, 
the most convenient choice is to take the z-axis parallel to {\bf q}. 
In fact, this choice leads to longitudinal 
and transverse asymmetries which correspond 
to the virtual Compton scattering asymmetries ${\cal A}_1$ and 
${\cal A}_2$, respectively ~\cite{Ans95,LP96}. 
\par
We recall that the quantity actually measured in the deep 
inelastic electron proton scattering experiments with a polarized target, 
is the longitudinal asymmetry corresponding to the choice 
of the z-axis parallel to the beam momentum ${\bf k}$. 
In this case the correspondence with the virtual Compton scattering 
asymmetries is more complex and ${\cal A}_z$ becomes a combination 
of ${\cal A}_1$ and ${\cal A}_2$. 
\par
Assuming ${\hat {\bf e}}_z // {\bf q}$, we obtain for 
${\cal A}_x$ and ${\cal A}_z$ 
%
%
\begin{eqnarray}
  {\cal A}_x &=& \rad2 ~\geff 
            ~~\frac{ \gve \ruptepupep \tGA G_E  
            + \gae \Big(G_E \tGM +G_M \tGE\Big) \rtepumep }
                 { \varepsilon ~G_E^2  +  \tau G_M^2  }
                    ~~~~, \nonumber  \\
             & &  \nonumber  \\
             & &  \label{e:Aptxz} \\
             & &  \nonumber \\
  {\cal A}_z &=&  2~\geff ~~\frac{  \gve \rtuptau \tGA G_M 
                          +  \gae G_M \tGM \tau \rumep2  }
                                 { \varepsilon ~G_E^2 +  \tau G_M^2 }
                  ~~~~.\nonumber  
\end{eqnarray}
\par
Of course, the asymmetry components when the z-axis is parallel to {\bf k} 
are related to the previous ones by a simple rotation 
in the scattering plane. In fact, if $\tkq$ is the angle between 
{\bf k} and {\bf q}, 
we have, with obvious notations 
%
%
\begin{eqnarray}
 {\cal A}_{x_k} &=& {\cal A}_{x_q} ~\cos\tkq - {\cal A}_{z_q} ~\sin\tkq 
                                            ~~~~, \nonumber  \\
 {\cal A}_{z_k} &=& {\cal A}_{x_q} ~\sin\tkq + {\cal A}_{z_q} ~\cos\tkq 
                                            ~~~~, \label{e:Axzkq} 
\end{eqnarray}
with
%
%
\begin{eqnarray}
 \sin\tkq &=& \dfrac{M}{E_e} 
              ~\sqrt{\dfrac{2 ~\varepsilon \tau}{1-\varepsilon}} 
                                                  ~~~~,   \nonumber \\
 & &                                            \label{e:cosinalfa} \\
 \cos\tkq &=& \Biggr(1+\dfrac{M}{E_e}\Biggr) 
              ~\sqrt{\dfrac{\tau}{1+\tau}}        ~~~~. \nonumber
\end{eqnarray}
\par
Coming back to Eq.(\ref{e:Aptxz}), several observations are in order. 
First, ${\cal A}_z$ does not depend on $\tGE$ and then on the proton 
strangeness form factor $G_E^{(s)}$. This follows from the fact that 
only the transverse components of the neutral currents contribute 
to ${\cal A}_z$. 
Second, the dependence of ${\cal A}_z$ 
on the proton axial form factor $\tGA$ is lowered with respect to that on 
the proton weak magnetic form factor $\tGM$ because $\gve \ll \gae$, in 
particular at backward angles where $\varepsilon \rightarrow 0$. 
This means that the longitudinal asymmetry ${\cal A}_z$ may be an useful 
quantity for an experimental determination of the proton strange magnetic 
moment, allowing for a measurement of $\mus$ complementary to that 
coming from the helicity asymmetry in the 
reaction ${\vec e}p \rightarrow e p$ as in the SAMPLE experiment. 
\par
In principle, this independence of $\tGE$ makes more convenient a 
measurement of $\tGM$ in the target asymmetry than in the helicity 
asymmetry of the ${\vec e}p \rightarrow ep$ reaction, $\Ae$, 
as it is clear recalling its expression \cite{MuDo94} 
%
%
\begin{eqnarray}
 \Ae  = - 2 ~\geff ~~\frac{ 
      \Big[ \gae \Big( \varepsilon ~G_E \tGE + \tau G_M \tGM  \Big) 
             +\gve \sqrt{\tau(1+\tau)(1-\varepsilon^2)} G_M \tGA  \Big] }
          { \varepsilon ~G_E^2 +  \tau G_M^2 }
      ~~. \label{e:Aep2}
\end{eqnarray}
\par
By the way, the longitudinal target asymmetry coincides, apart from the 
sign, with the helicity asymmetry for backward scattering 
($\vartheta_{e^\prime} = 180^\circ \Rightarrow \varepsilon = 0$) 
as a consequence of the helicity conservation in the ERL for the electrons. 
\par
A final remark following from the comparison of $\Ae$ 
with both the components of the target asymmetry, concerns the different 
sensitivity on $\tGA$ due to the different dependence on $\varepsilon$. 
In fact, unlike $\Ae$, the terms containing the weak vector 
form factors in the target asymmetries, usually dominant because 
$\gae \gg \gve$, are suppressed at forward angles 
($\varepsilon \rightarrow 1$)
by the kinematic factors making substantial the impact of $\tGA$. 
\par
Therefore, a determination of $\tGA$ alternative to 
that deriving from $\nu/{\bar \nu}$ scattering experiments could be 
carried out in this kind of PV electron scattering experiments. However, 
as in actual experiments the scattering angle can not be less than 
$\sim$ 10$^\circ$, the suppression mentioned above is only partial and thus 
a good knowledge of the weak vector form factors is a prerequisite of 
such an experiment. 
\subsection{Nucleon electromagnetic and weak form factors}
From the structure of the em and weak vector current operators in terms 
of the SU(3)-singlet and -octet currents it follows that the nucleon 
weak vector form factors are given by
%
%
\begin{eqnarray}
  {\tilde G}_{E,M}(Q^2) &=& \dfrac{1}{2} \xi_V^{T=1} G_{E,M}^V(Q^2) \tau_3 
                  + \dfrac{\sqrt{3}}{2} \xi_V^{T=0} G_{E,M}^S(Q^2) 
                                        \nonumber \\
                 &+&                    \xi_V^{(0)} G_{E,M}^{(s)}(Q^2) 
                 ~~~~,\label{e:GEMW}
\end{eqnarray}
with  $\tau_3$=+1,~-1 for the proton and neutron, respectively. 
~$G_{E,M}^{S(V)}$ is the isoscalar (isovector) combination of the em 
Sachs form factors, $G^{(s)}_{E,M}$ is the strange-quark contribution and 
the couplings are appropriate linear combinations of quark weak vector 
charges. In the Standard Model they have the values
%
%
\begin{equation}
 \xi^{T=1}_V = 2(1-2 \sin^2 \vartheta_W) ~~~~,~~~~
 \sqrt{3} \xi^{T=0}_V = -4 \sin^2 \vartheta_W ~~~~,~~~~ 
 \xi^{(0)}_V = - 1
                  ~~~~~.\label{e:csivector}
\end{equation}
\par
The nucleon has not net strangeness, so that $G^{(s)}_E(0)=0$. This is 
the only theoretical constraint about the strangeness form factors, 
which, according to ~\cite{MuDo94}, we take in the form 
%
%
\begin{eqnarray}
  \GES &=&  - ~\dfrac{1}{6} ~r^{2~[S]}_s ~Q^2
               ~G_D^V(Q^2) ~(1+\laes \tau)^{-1}
                                               ~~~~~, \nonumber \\
                 & &                  \label{e:GEMstrani} \\
  \GMS &=&   \mus  ~G_D^V(Q^2) ~(1+\lams \tau)^{-1}
                                               ~~~~~, \nonumber 
\end{eqnarray}
i.e. an extension of the Galster parametrization ~\cite{Galster71}~ 
commonly used for the nucleon em form factors
%
%
\begin{eqnarray}
   G_E^p(Q^2) &=& G_D^V(Q^2)~~~~~,~~~~~
   G_E^n(Q^2) = -\mu_n \tau G_D^V(Q^2) (1+5.6 \tau)^{-1} ~~~~~\nonumber \\
         & &                       \label{e:Galster} \\
   G_M^p(Q^2) &=&  \mu_p  G_D^V(Q^2)~~~~~,~~~~~
   G_M^n(Q^2) =  \mu_n  G_D^V(Q^2)  ~~~~~,\nonumber
\end{eqnarray}
where ~$G_D^V(Q^2) = (1+Q^2/M_V^2)^{-2}$ , 
with a cut-off mass squared $M_V^2$=0.71 $GeV^2$. 
\par
The Sachs strangeness radius $r^{2~[S]}_s$, 
which characterizes the low $Q^2$ behaviour of $G^{(s)}_E$, is related to 
the more familiar Dirac strangeness radius 
%
%
\begin{equation}
  \rdues \equiv - 6 \left [ \dfrac{d~F_1^{(s)}(Q^2)}{d~Q^2} \right ]_{Q^2=0} 
                                      ~~~~~~~~~,\label{e:r2s}
\end{equation}
by the relation 
%
%
\begin{equation}
   r^{2~[S]}_s  = r^2_s  + ~\dfrac{3}{2 M^2} ~\mu_s                   
                                     ~~~~~. \label{e:r2sSACDIR}  
\end{equation}
\par
Very little is known about the values of $\mus$ and $\rdues$ even if many 
calculations of the strangeness vector form factors have been carried out 
using different approaches (lattice calculations, effective Lagrangian, 
dispersion relations, hadronic models). 
The predictions of the strangeness moments are quite different in different 
approaches and can also largely vary within a given approach because of the 
need of additional assumptions and approximations. 
In particular, $\rdues$ is predicted to be positive in the dispersion 
theory analysis of the nucleon isoscalar form factors 
~\cite{Jaffe89,Hammer96}, 
of the same order of magnitude but negative by the 
chiral quark-soliton model ~\cite{Kim95}~ and negative but of 
two order of magnitude smaller by the kaon-loop 
calculations ~\cite{MuBu94}. 
A negative value of $\rdues$ is also preferred by the analysis 
\cite{Garvey93}~ of the $\nu p/{\bar \nu} p$ elastic scattering data 
\cite{Ahrens87}~  which, however, has been criticized for the use 
of a unique cut-off mass for the three SU(3) axial-vector form factors. 
\par 
The different existing models widely disagree also about sign and magnitude 
of $\mus$ which is predicted to range from 
$\mus=0.4 ~\mu_N$ ~\cite{HongPark93}~ in the chiral hyperbag model 
to $\mus=-0.75 ~\mu_N$ ~\cite{Leinw96}~ using QCD equalities 
among the octet baryon magnetic moments. 
\par
Analogously to (\ref{e:GEMW}), the axial-vector form factor can be decomposed 
in terms of the 3rd and 8th SU(3) octet components and of the possible 
strange component 
%
%
\begin{equation}
  {\tilde G}_A(Q^2) = \dfrac{1}{2} ~\xi_A^{T=1} ~G_A^{(3)}(Q^2) ~\tau_3 
                 + \dfrac{1}{2} ~\xi_A^{T=0} ~G_A^{(8)}(Q^2) 
                 + \xi_A^{(0)} ~G_A^{(s)}(Q^2) 
                 ~~~~, \label{e:GAP} 
\end{equation}
with coupling constants dictated at the tree level by the quark axial charges 
%
%
\begin{equation}
 \xi^{T=1}_A = - 2 ~~~~,~~~~ 
 \xi^{T=0}_A = 0 ~~~~,~~~~ 
 \xi^{(0)}_A = 1
                  ~~~~~.\label{e:csiaxial}
\end{equation}
\par
Note that in this limit the isoscalar component of ${\tilde G}_A$ fully comes 
from the strange quark contribution. Information on the $Q^2=0$ value of the 
SU(3) octet form factors derives from charged current weak interactions. 
From neutron $\beta$-decay and strong isospin symmetry it follows 
$G^{(3)}_A(0) = (D+F) \equiv g_A = 1.2601 \pm 0.0025$ ~\cite{PDG96}, 
while from hyperon $\beta$-decays and flavor SU(3) symmetry it follows 
$G^{(8)}_A(0) = (1/\sqrt{3}) (3F-D) = 0.334 \pm 0.014$ ~\cite{Close93}, 
%
%
$D$ and $F$ being the 
associated SU(3) reduced matrix elements. 
The $Q^2$ dependence of these form factors can be adequately parametrized 
with a dipole form 
%
%
\begin{equation}
    G_D^A(Q^2) = (1+Q^2/M_A^2)^{-2}  ~~~~~, \label{e:GAdip}
\end{equation}
with a cut-off mass $M_A$=1.032 GeV. 
The same dipole form is suggested in \cite{MuDo94}~ 
for the strange axial vector form factor 
%
%
\begin{equation}
  \GAS =  \gaesse ~G_D^A(Q^2) ~(1+\lambda_A^{(s)} \tau)^{-1}
                                 ~~~~~.  \label{e:GAstrano}
\end{equation}
Here again, lacking theoretical constraints on $G^{(s)}_A(0)$ and 
because of the model dependence of the theoretical estimates, 
values of $\gaesse$  have to be extracted from the 
experiments and the first indications came from the 
BNL $\nu p / {\bar \nu} p$ experiment~\cite{Ahrens87}~ and from the EMC 
data~\cite{Ashman8889}. 
\par
As for the weak coupling constants we emphasize that the values 
(\ref{e:csivector}) and (\ref{e:csiaxial}) are those predicted by 
the ewk Standard Model at the tree-level.
In a realistic evaluation of the amplitude of any electron-hadron process 
one has to consider the radiative corrections to these values. 
Such corrections $R_{V,A}$, 
amounting to a factor (1+$R_{V,A}$) in all the coupling 
constants except in $\xi^{T=0}_A$ which becomes $\sqrt{3} R^{T=0}_A$, 
are very difficult to calculate because they receive 
contributions from a variety of processes (higher-order terms in ewk 
theory, hadronic physics effects,...).  They have been estimated by 
various authors (for a review see Ref.\cite{MuDo94}~ and citations therein) 
using different approaches and approximations with results in qualitative 
agreement. 
More precisely, $R_V$ are estimated to be of the order of a few percent 
and $R_A$ of the order of some tenth of percent. Therefore, while 
$R_V$ can be neglected, the radiative corrections $R_A$ must, 
in principle, be taken into account. 
\section{Results}
In this paper we report our results on the polarized target asymmetry in 
the momentum transfer region $Q^2 \le$ 1 (GeV/c$^2$). 
As for the parameters which determine the $Q^2$ fall-off of the expressions 
(\ref{e:GEMstrani},\ref{e:GAstrano}) of the strange form factors, which are 
completely unknown, we take the values $\laes=5.6, ~~\lams=0, ~~\laas=0$. 
While the value for $\laes$ is suggested by the analogy with the Galster 
parametrization of $G_E^n$, the choice for $\lams$ and $\laas$ seems to us 
rather reasonable as we do not consider too high $Q^2$ values. 
\par
Furthermore, as reference values for the axial radiative corrections we 
adopt $\RATZERO=-0.62$ and $\RAT1=-0.34$ given by Musolf and Holstein 
\cite{MuHo90}~ using for the hadronic contributions the so-called 
best estimates for the weak meson-nucleon vertices of Ref.\cite{Despla80}. 
On the contrary, lacking a reliable estimate of the radiative correction 
to the strangeness axial coupling constant, we take $\RAS=0$. 
Finally, we use $\gaesse=-0.15$ as reference value of the strangeness axial 
charge, as deduced from the neutrino scattering experiment~\cite{Ahrens87}. 
\begin{figure}
\centerline{\psfig{figure=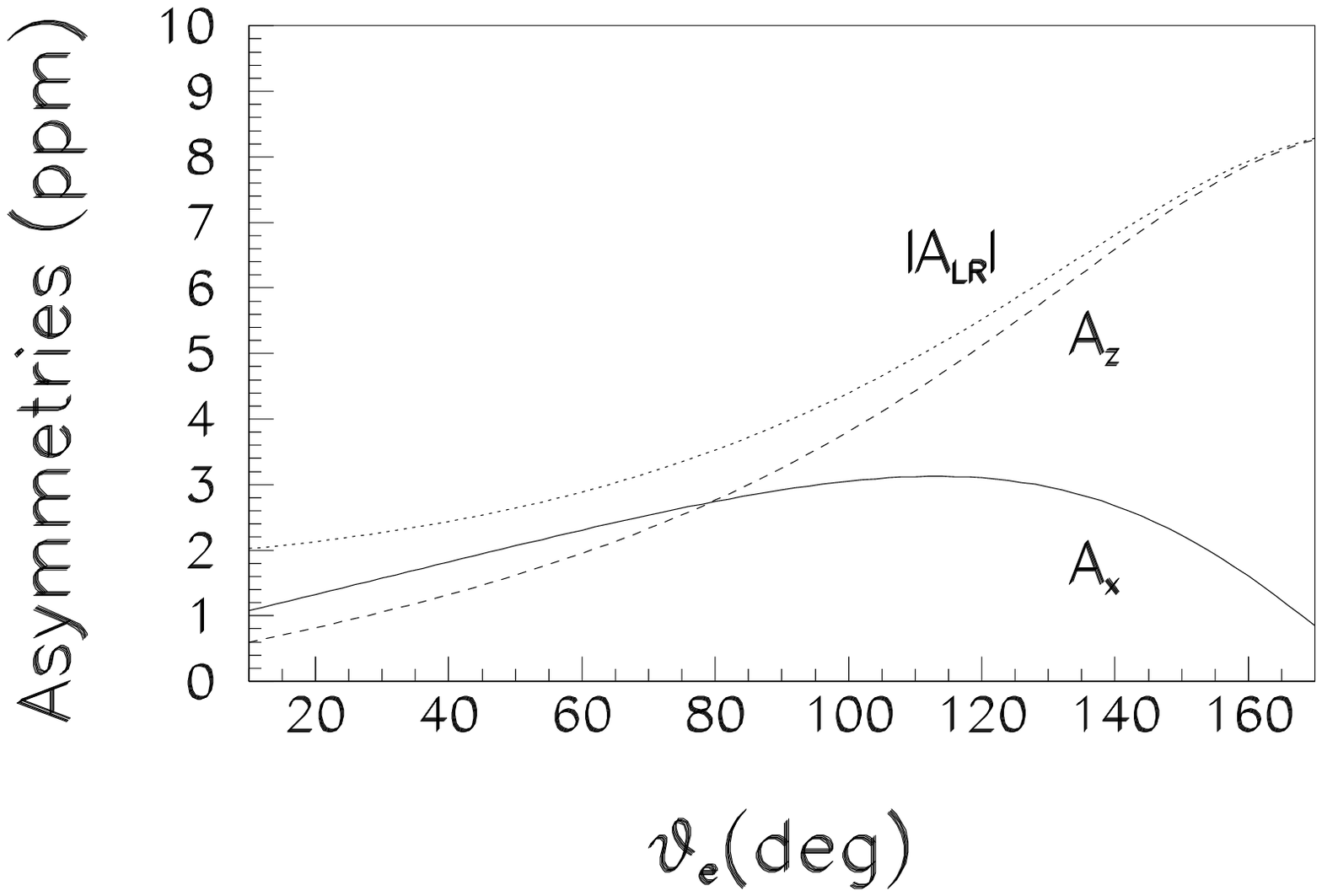,width=14.0cm}} 
\caption{
   Angular distribution of the target asymmetry $\Ax(\tep)$ (full line), 
   $\Az(\tep)$ (dashed line) and of the modulus of the helicity asymmetry 
   $\Ae(\tep)$ (dotted line) at $Q^2$=0.1$(GeV/c)^2$, with $\mus$=0.23~$\mu_N$ 
   \cite{Mue97} and $\rdues$=0.16~fm$^2$ \cite{Jaffe89}. 
}
\end{figure}
\par
Let us start considering the angular distribution of the polarized 
target asymmetries. 
In Fig.1 we plot  $\Ax(\tep)$ and $\Az(\tep)$ for $Q^2$=0.1 (GeV/c)$^2$, 
calculated with Jaffe's value \cite{Jaffe89}~ of the strangeness radius 
($\rdues$=0.16~fm$^2$) and with the central experimental value of the 
strange magnetic moment ($\mus=0.23 ~\mu_N$) \cite{Mue97}. 
The asymmetries, which are quite small at forward angles, show 
a remarkably different angular dependence as $\tep$ increases. 
In fact, while $\Ax(\tep)$ increases very slowly, reaches a maximum for 
$\tep \simeq$ 110$^\circ$ and then decreases at backward angles, $\Az(\tep)$ 
increases continuously with $\tep$ reaching at backward angles values 
which are one order of magnitude greater than those at forward angles. 
\par
The angular behavior of the target asymmetries can be easily understood 
recalling their expression (\ref{e:Aptxz}). In fact, at forward angles 
$\varepsilon \simeq 1$ and the surviving term in both the asymmetries is 
that proportional to $\gve$. On the contrary, at backward angles 
$\varepsilon \rightarrow 0$ and $\Ax(\tep) \rightarrow 0$ while 
$\Az(\tep)$ reaches its maximum. 
\begin{figure}
\centerline{\psfig{figure=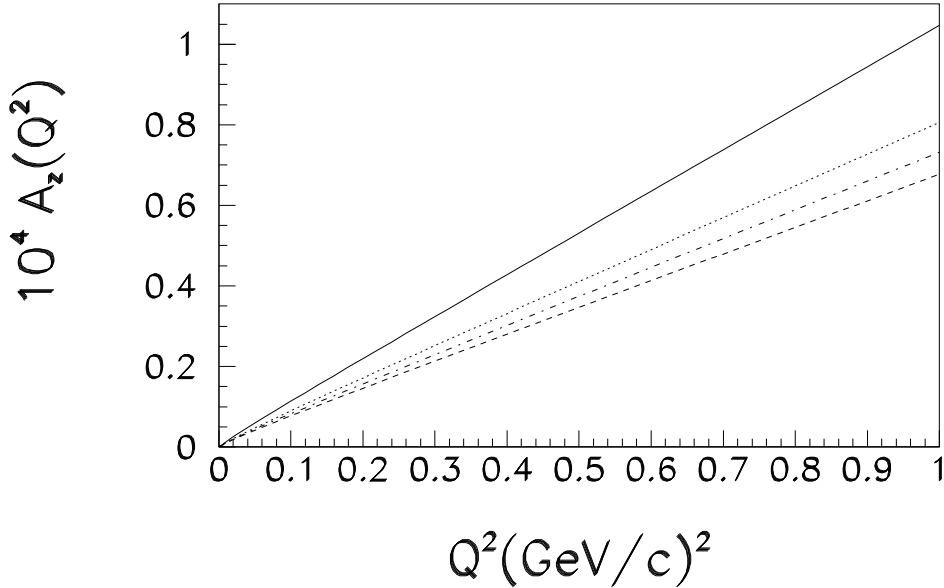,width=14.0cm}} 
\caption{
  Dependence of the target asymmetry $\Az(Q^2)$ on the strange magnetic 
  moment $\mus$, at $\tep=170^\circ$ and for 
  $\rdues$=0.16 ~fm$^2$ ~\cite{Jaffe89}. 
  The solid line is for $\mus$=-0.75~$\mu_N$ ~\cite{Leinw96}, 
  the dashed line for $\mus$=0.40~$\mu_N$ ~\cite{Hammer96},
  the dotted line for $\mus$=0 and the dot-dashed line for 
  $\mus$=0.23~$\mu_N$ ~\cite{Mue97}. 
}
\end{figure}
\par
For comparison, we have also drawn in Fig.1 the modulus of the helicity 
asymmetry $\Ae(\tep)$ which has an angular dependence very similar to that of 
$\Az(\tep)$. In fact, $|\Ae|$ which is slightly greater than $\Az$ at forward 
angles, converges to $\Az$ for $\tep \rightarrow 180^\circ$, as discussed in 
the previous section. There we pointed out that the longitudinal target 
asymmetry at backward angles may be an useful quantity for an experimental 
determination of $\mus$. 
In order to show the effect on  $\Az$ of variations in the strangeness 
magnetic moment we report in Fig.2 our results at $\tep$=170$^\circ$ 
for a selected set of predictions of $\mus$. Among the values given by the 
different models we have chosen those defining the theoretical range of 
$\mus$, i.e. $\mus=-0.75~\mu_N$ \cite{Leinw96}~ and 
$\mus=0.4~\mu_N$ \cite{HongPark93}. Also reported are the curves 
corresponding to the experimental value of $\mus$ \cite{Mue97}~ and to 
$\mus=0$. Note that the Dirac strangeness radius has been held fixed, 
$\rdues=0.16$~fm$^2$ as deduced by Jaffe. 
This comparison makes evident the sensitivity on $\mus$ of the target 
asymmetry for electrons scattered in the backward direction, 
sensitivity which is very strong for $Q^2$=1 (GeV/c)$^2$ but already 
noticeable even for lower $Q^2$ values. 
\begin{figure}
\centerline{\psfig{figure=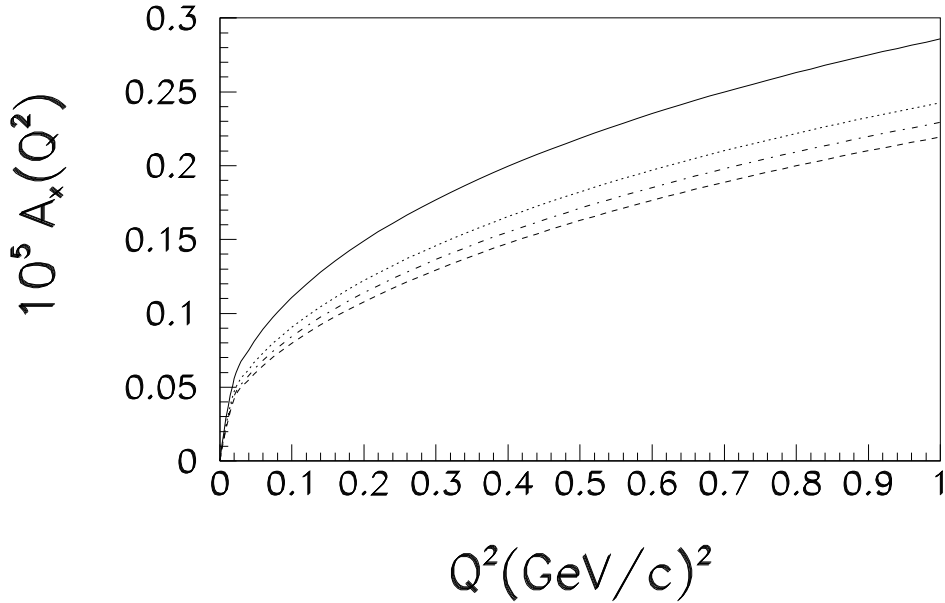,width=14.0cm}} 
\caption{
  The same as in Fig.2 for the target asymmetry $\Ax(Q^2)$. 
}
\end{figure}
\par
Of course, because at backward angles $\Az$ and $\vert\Ae\vert$ are nearly 
coincident, we can repeat for $\Az$ all the considerations made for 
$\Ae$ on the precision reachable in an extraction of $\mus$ from a 
measurement of $\Ae$. For example, at $Q^2$=0.1 (GeV/c)$^2$, 
the uncertainties affecting the other quantities determining the asymmetries 
(radiative corrections to the coupling constants, uncertainties in $\tGA$ 
arising from $\gaesse$ and $M_A$, ... ) make it necessary a measurement 
within a 10$\%$ error to extract $\mus$ with a limit 
$\vert \delta\mus \vert \approx 0.22$ \cite{MuDo94}. 
In the previous section we pointed out that the longitudinal target 
asymmetry has the important characteristic of being independent of 
$\tGE$ and then, in particular, of $G_E^{(s)}$. 
Unfortunately, the positive consequences of this fact, 
are made less significant by all the above mentioned uncertainties on 
the other parameters. 
\par
For completeness, we report in Fig.3 the transverse target asymmetry $\Ax$, 
in the same kinematical conditions as before and for the same set of 
values of $\mus$. 
Even if $\Ax$ shows a good sensitivity to $\mus$, it is nearly an order of 
magnitude smaller than $\Az$ and then less suitable for an 
experimental measurement. 
\begin{figure}
\centerline{\psfig{figure=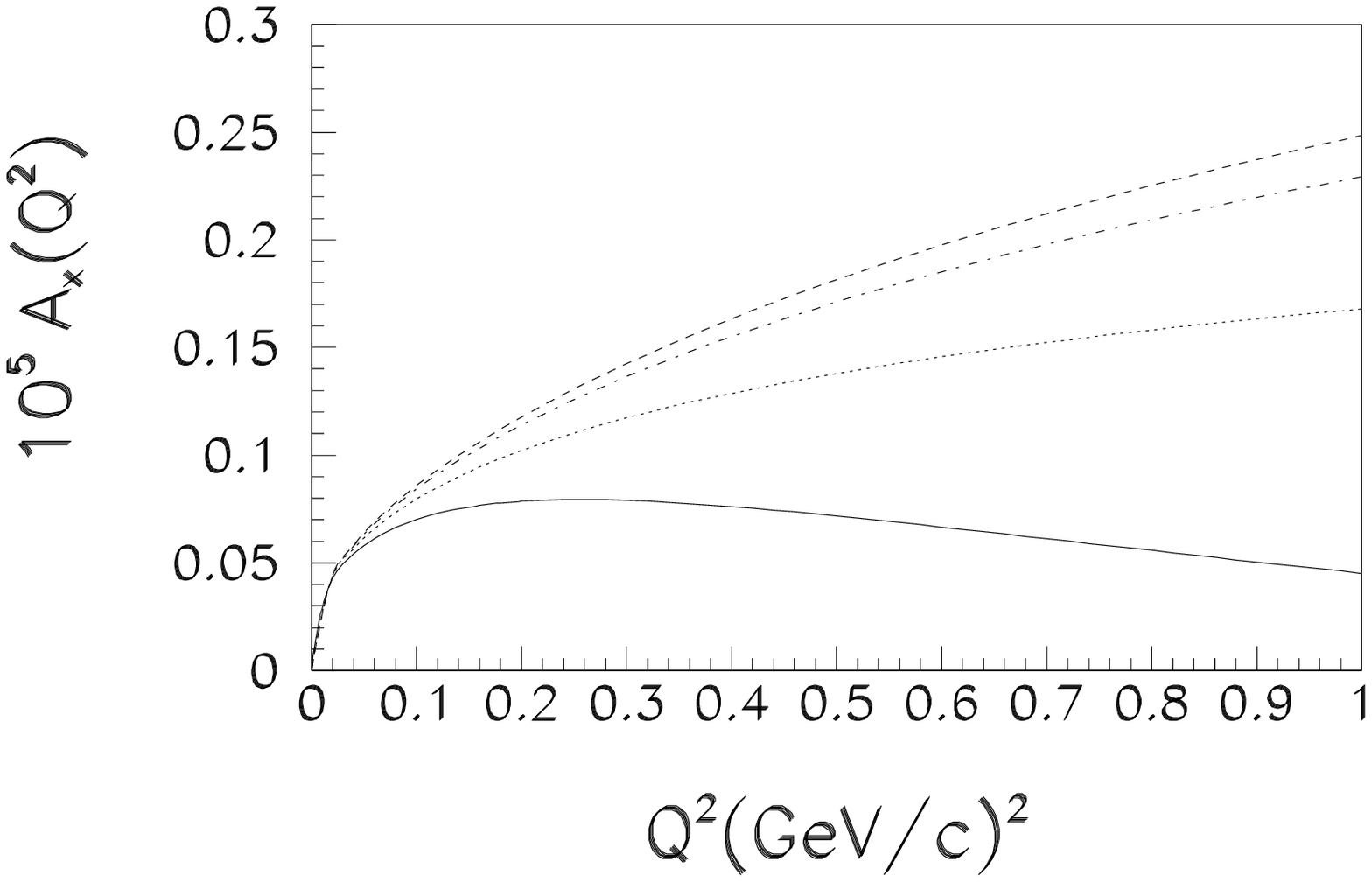,width=14.0cm}} 
\caption{
  Dependence of the target asymmetry $\Ax(Q^2)$ on the strangeness radius 
  $\rdues$, at $\tep=170^\circ$ and for $\mus$=0.23~$\mu_N$ ~\cite{Mue97}. 
  The solid line is for $\rdues$=-0.32 ~fm$^2$ ~\cite{Kim95},
  the dashed line for $\rdues$=0.21 ~fm$^2$ ~\cite{Hammer96}, 
  the dotted line for $\rdues$=0 and the dot-dashed line for 
  $\rdues$=0.16 ~fm$^2$ ~\cite{Jaffe89}. 
}
\end{figure}
\par
On the contrary, $\Ax$ could be an useful quantity to constrain $\rdues$. 
According to Eq.(\ref{e:Aptxz}), in order to minimize the impact of the 
poorly known $\tGA$, it is convenient to consider backward scattered 
electrons.  
The sensitivity of $\Ax$ to the strangeness radius is shown in Fig.4, 
where we compare our results of $\Ax$ for $\tep=170^\circ$, having fixed the 
proton strangeness magnetic moment to its experimental value and 
for a restricted selection of predicted $\rdues$. 
Besides that given by Jaffe and $\rdues=0$, we have used the two almost 
opposite values $\rdues=0.21$~fm$^2$, deduced by Hammer et al. 
\cite{Hammer96}~ in their revised dispersion analysis and 
$\rdues=-0.32$~fm$^2$ obtained by Kim et al. \cite{Kim95}~ in a 
chiral quark soliton model. 
\par
At first sight, a measurement of $\Ax$ in the backward direction 
could lead to discriminate between the different models. 
Actually, the precision on the extraction of $\rdues$ from such experiment 
is strongly limited by the error induced by the uncertainty in the other 
quantities determining $\Ax$ and particularly in $\mus$. 
To be more quantitative on the precision reachable in a determination 
of $\rdues$, let us consider again the expression of $\Ax$, 
Eq.(\ref{e:Aptxz}), and write it in the form 
%
%
\begin{equation}
 \Ax = \Ax^0 \Big(1 + a_x ~\rdues + b_x ~\mus 
                    + c_x ~\RAT1 + d_x ~\RATZERO + e_x ~\gaesse \Big) 
                      ~~~~,\label{e:abcdex}
\end{equation}
which exibits the dependence on the strangeness radius, 
magnetic moment, axial charge and on the radiative corrections to 
the axial-vector coupling constants. 
The fractional change induced in the transverse asymmetry is given by 
%
%
\begin{equation}
 \dfrac{\delta\Ax}{\Ax} \simeq
               \Big(a_x ~\delta\rdues + b_x ~\delta\mus + c_x ~\delta\RAT1 
                   + d_x ~\delta\RATZERO  + e_x ~\delta\gaesse \Big) 
                      ~~~~.\label{e:dAsuAx}
\end{equation}
\par
The values of the coefficients in Eq.(\ref{e:abcdex}) and (\ref{e:dAsuAx}) 
for $\tep=170^\circ$ and  $Q^2=1 (GeV/c)^2$ are 
$a_x$=0.208~fm$^{-2}$ , $b_x$=-0.312 , $c_x$=0.172 , 
 $d_x$=-0.048 , $e_x$=-0.136, 
and it turns out that the impact of $\delta\mus$ , $\delta\RAT1$ and 
$\delta\gaesse$ on a determination of $\rdues$ is strong. 
In fact, even with an 
experimental error of 10$\%$ in the measurement of $\Ax$, assuming 
$\delta\RAT1 = \pm 0.28$ as estimated by Musolf and Holstein \cite{MuHo90}, 
$\delta\gaesse = \pm 0.09$ as given by neutrino scattering 
experiments \cite{Ahrens87}, and assuming to know $\mus$ within $\pm$0.1, 
one can not get for $\rdues$ a precision better than 
$\vert \delta\rdues \vert \approx 0.6$. It is to be concluded that 
the transverse asymmetry compares only slightly favorably with the beam 
helicity asymmetry as for the determination of $\rdues$. 
\par
Finally, we consider the possibilities provided by a measurement of target 
asymmetries for a determination of $\gaesse$. In Fig.5  we plot 
$\Ax$ and $\Az$ at forward angle ($\tep$=10$^\circ$) as a function of $Q^2$ and 
for different values of $\gaesse$ within the range defined by neutrino 
experiments \cite{Ahrens87}. 
The other parameters are taken at their reference values. The variations 
induced by $\gaesse$ are very similar in $\Ax$ and $\Az$, slowly increasing 
with $Q^2$ and reach about 10$\%$ at $Q^2$=1 (GeV/c)$^2$. 
To give a quantitative estimate of the error with which $\gaesse$ could 
be extracted from a PV target asymmetry measurement, we limit our 
consideration to $\Az$ which does not take contribution from $\tGE$. 
Therefore, analogously to Eq.(\ref{e:dAsuAx}), we have 
%
%
\begin{equation}
 \dfrac{\delta\Az}{\Az} \simeq
               \Big( b_z ~\delta\mus + c_z ~\delta\RAT1 
                   + d_z ~\delta\RATZERO  + e_z ~\delta\gaesse \Big) 
                      ~~~~,\label{e:dAsuAz}
\end{equation}
and from the evaluation at $Q^2$=1 (GeV/c)$^2$ and $\tep$=10$^\circ$, we obtain 
$b_z$=-0.243 , $c_z$=0.473 , $d_z$=-0.133 , $e_z$=-0.375 .
\par
Clearly, the largest effects on $\Az$ come from the uncertainties in 
$\RAT1$ and $\gaesse$, but also $\mus$ and $\RATZERO$ have a sizeable 
impact. Therefore, a good knowledge of the radiative corrections and 
a precise determination of $\mus$ are requested in order that a 
measurement of $\Az$ could be exploited for constraining the value 
of $\gaesse$. 
\begin{figure}
\centerline{\psfig{figure=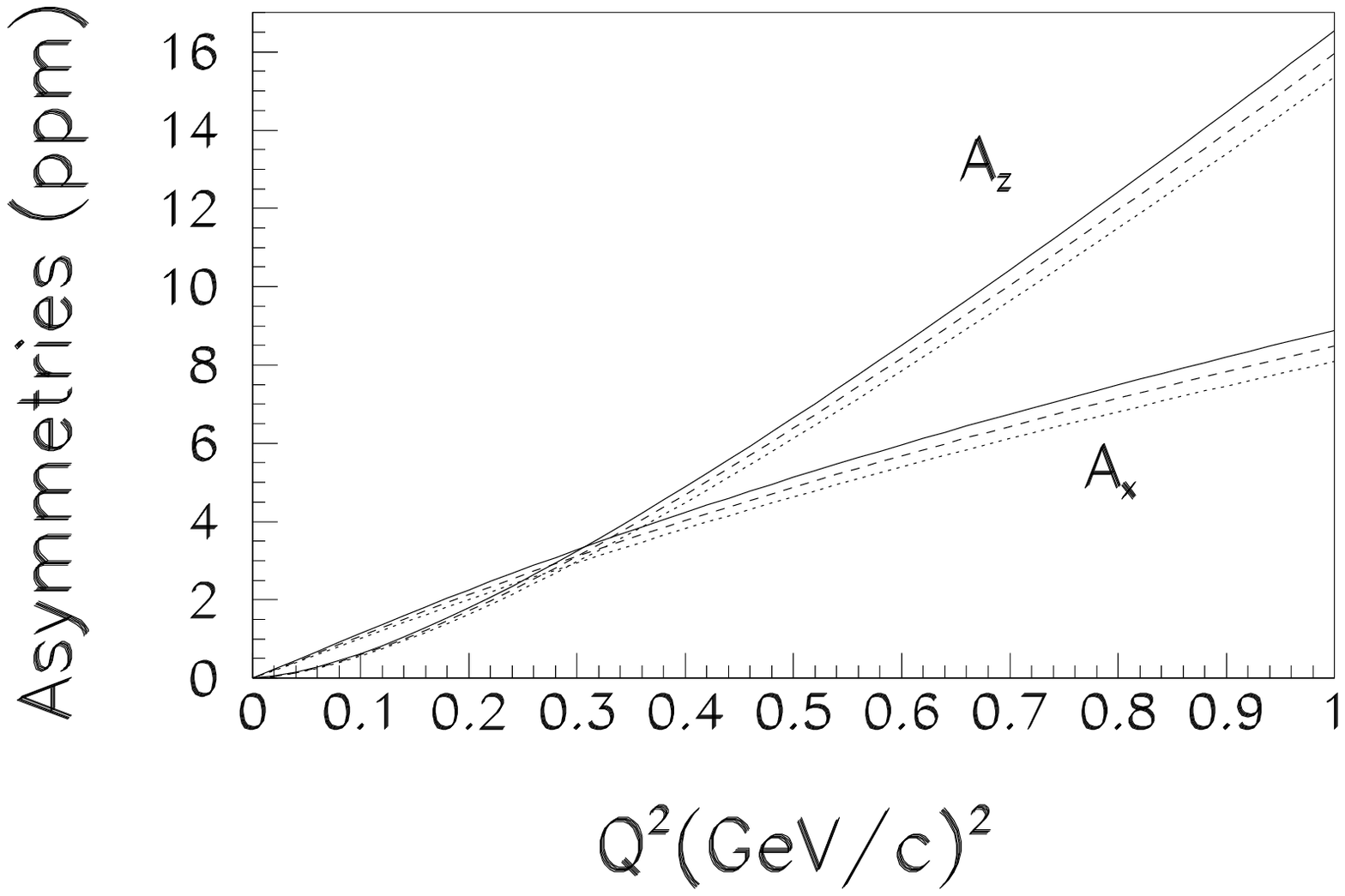,width=14.0cm}} 
\caption{
  Dependence of the target asymmetry $\Ax(Q^2)$ and $\Az(Q^2)$ 
  on the axial charge $\gaesse$, at $\tep=10^\circ$, 
  for $\mus$=0.23~$\mu_N$ ~\cite{Mue97} and $\rdues$=0.16~fm$^2$ ~\cite{Jaffe89}. 
  The dashed line is for the central value $\gaesse$=-0.15 while 
  the full line and dotted line are for $\gaesse$=-0.24 and $\gaesse$=-0.06, 
  respectively.
}
\end{figure}
\section{Conclusions}
\par
The aim of this paper was to extend the possible PV observables which could 
be used for an experimental determination of the weak form factors of the 
proton. To this end we have considered the asymmetries of the elastic 
electron-proton scattering cross section arising from the polarization 
of the proton target. 
\par
Concerning such an experiment, it seems to us that it should become 
feasible in the near future because the tecnique of polarized target 
production has so much improved that it is nowadays possible to obtain 
hydrogen targets highly polarized, with an arbitrary direction of 
polarization, and with fast and efficient spin reversal. 
\par
First of all, we have carefully rederived the expression of the cross 
section in the electroweak theory valid for an arbitrary target 
polarization. Indeed, such a process was already addressed in the past 
by several authors but with incorrect results. 
\par
Then, we have decomposed the PV analyzing power, which is a vector lying 
in the scattering plane, considering the proton polarization along and 
perpendicular to the momentum transfer {\bf q}. This is indeed the most 
convenient decomposition from the theoretical point of view, leading to a 
longitudinal asymmetry independent of the electric weak form factor.  
For comparison, we recall that the helicity asymmetry of the ${\vec e}-p$ 
scattering depends also on $\tGE$. 
\par
However, the actual calculations of $\Az$ for backward angles, which is the 
most suitable kinematical situation for an experimental determination of 
$\mus$, lead to an expected precision on $\mus$ which is of the same order 
as the one obtained in the helicity asymmetry experiment.
\par
We have also shown that the transverse target asymmetry $\Ax$ is rather 
sensitive to the strangeness radius $\rdues$ in the case of backward 
detected electrons. However, it turns out from the calculations that the 
precision in the extraction of $\rdues$ is strongly limited by the 
uncertainties in the other quantities (particularly in $\mus$) affecting 
$\Ax$. 
\par
Another peculiarity of the target asymmetries $\Ax$ and $\Az$ with respect 
to the beam asymmetry, is that their dependence on the axial form factor 
can be enhanced over that on the vector form factors. In fact, in the 
strict forward scattering ($\tep$ = 0$^\circ$) with the electron kinematical 
parameter $\varepsilon$=1, the target asymmetries are determined by 
$\tGA$ only. In practice, the finite dimensions of the detectors prevent 
too small scattering angles so that the influence of $\tGE$ and $\tGM$ 
can be greatly reduced but non completely cancelled. Anyway, a 
measurement of the target asymmetries in this kinematics could be 
useful for independently constraining the value of $\GAS$ given by the 
neutrino scattering experiments. 
\section{Acknowledgements}
\par
This work was partly supported by Ministero della Universit\`a e 
della Ricerca Scientifica of Italy.
\newpage
\end{document}